\documentclass[12pt]{article}
\usepackage{a4wide, amsfonts}

\newcommand{\R}{\hbox{\upright\rlap{I}\kern 1.7pt R}}

\newcommand{\news}{\setcounter{equation}{0}}
\font\upright=cmu10 scaled\magstep1
\newcommand{\g}{{\cal G}}
\newcommand{\tr}{{\rm Tr}}
\newcommand{\de}{\delta}
\newcommand{\1}{\hbox{\upright\rlap{I}\kern 1.2pt 1}}
\begin{document}

\thispagestyle{empty}

\title{\vskip -70pt
  \begin{flushright}
    {\normalsize DAMTP 98-10. \\ physics/9802012.} 
  \end{flushright}
  \vskip 15pt
  {\bf \Large \bf Invariant tensors and Casimir operators for simple
  compact Lie groups}
  \vskip 10pt
}

\author{Arthur J. Mountain\thanks{E-mail : A.J.Mountain@damtp.cam.ac.uk} \\
  {\normalsize {\sl Department of Applied Mathematics
      and Theoretical Physics,}}\\
  {\normalsize {\sl University of Cambridge, Silver Street,}} \\
  {\normalsize {\sl Cambridge CB3 9EW, United Kingdom.}}
  }

\date{March 1998}

\maketitle

\vskip 100pt

\begin{abstract} 
  \noindent The Casimir operators of a Lie algebra are in
  one-to-one correspondence with the symmetric invariant tensors of
  the algebra. There is an infinite family of Casimir operators whose
  members are expressible in terms of a number of primitive Casimirs
  equal to the rank of the underlying group. A systematic derivation
  is presented of a complete set of identities expressing
  non-primitive symmetric tensors in terms of primitive tensors.
  Several examples are given including an application to an
  exceptional Lie algebra.
\end{abstract}

\newpage
\baselineskip17pt
\section{Introduction}
\news 
The Casimir operators for a simple Lie algebra are constructed from
symmetric tensors invariant under the adjoint action of the group.
Given a simple compact Lie group of rank $l$, it is known that there
are $l$ such primitive symmetric invariant tensors of order $m_i, i=1
\ldots l$. The orders $m_i$ are in fact also known because of the
correspondence of the primitive symmetric invariant tensors with
cocycles of order $(2m_i-1)$ which are representatives of the de Rham
cohomology classes of $G$. However, most methods of constructing
symmetric invariant tensors give rise naturally to an infinite family
of them, of which only the finite number $l$ can be primitive.
\vskip5pt
\noindent In this paper, we consider the set of invariant tensors
arising from symmetrized traces of products of Lie algebra basis
elements. These correspond to Casimir operators and arise naturally in
many physical theories. It is often useful to be able to express this
(infinite) family of tensors in terms of the $l$ primitive tensors
(see above). We present a method which systematically generates a
complete set of identities expressing non-primitive symmetric
invariant tensors of this form in terms of primitive tensors. We
illustrate our methods for the classical series of Lie algebras and perform
explicit calculations for the examples $A_2$, $B_2$, and $G_2$. 

\subsection*{Notation and conventions}
In this paper, we will follow the ``physical'' convention for Lie
groups and use hermitian generators $\{X_i\}$ and the normalization
$\tr (X_i X_j)=2\de_{ij}$. When the generators are assumed to be in
{\it matrix} form, we take them to be in the defining representation
of the algebra. We will assume there is no distinction between upper
and lower indices, their position being dictated by notational
convenience. We will take $\epsilon_{i_1i_2\ldots i_n}$ to be the
standard totally antisymmetric tensor, with $\epsilon_{123\ldots
  n}=1$. We will define from this the unit-weight symbol

\begin{equation}
  \epsilon^{j_1j_2\ldots j_n}_{i_1i_2\ldots i_n} = \frac{1}{n!}
    \epsilon_{i_1i_2\ldots i_n} \epsilon^{j_1j_2\ldots j_n} =
    \frac{1}{n!} \sum_{\sigma \in S_n}(-1)^{\pi(\sigma)}
    \delta^{j_{\sigma(1)}}_{i_1} \ldots \delta^{j_{\sigma(n)}}_{i_n},
\end{equation}
where $\pi(\sigma)$ is the parity of the permutation $\sigma$.

\section{Invariant symmetric tensors and Casimirs}
\news
Let $\g$ be a simple algebra of rank $l$ with basis $\{X_{i}\}$,
$[X_i,X_j]=f_{ij}^k X_k$, $i=1\dots\mbox{dim}\,\g$, and let $G$ be its
(compact) associated Lie group.  Let $\{\omega^{j}\}$ be the dual
basis in $\g^*$, $\omega^{j}(X_i)=\delta_i^j$, and consider a
$G$-invariant symmetric tensor $h$ of order $m$

\begin{equation}
  h=h_{i_1\dots i_m} \omega^{i_1}\otimes\dots\otimes\omega^{i_m} \quad.
\end{equation}
The $G$-invariance (ad-invariance) of $h$ means that

\begin{equation}
  \sum_{s=1}^m f^\rho_{\nu i_s} h_{i_1\dots \widehat{i_s}\rho i_{s+1}
    \dots i_m}=0 \quad.
  \label{invcond}
\end{equation}
This is the case for the symmetric tensors $k^{(m)}$ given by the 
coordinates

\begin{equation}
  k^{(m)}_{i_1\dots i_m}= {\rm sTr}(X_{i_1}\dots X_{i_m}) \equiv
    \tr(X_{(i_1}\dots X_{i_m)})\quad.
  \label{symmtrace}
\end{equation}
where ${\rm sTr}$ is the unit-weight symmetric trace, $\displaystyle
{\rm sTr} (X_{i_1}\dots X_{i_m})= \frac{1}{m!} \sum_{\sigma\in S_m}
\mbox{Tr}(X_{i_{\sigma(1)}}\dots X_{i_{\sigma(m)}})$. Indices inside
round brackets $(i_1,\dots,i_m)$ will also be understood as
symmetrized with unit weight. In fact \cite{invGR}, a complete set of
$l$ primitive (see below) invariant tensors may be constructed in this
way by selecting suitable representations.  
\vskip5pt
\noindent Let $G$ be compact so that the Killing tensor may be taken as the unit
matrix and let $h_{i_1\dots i_m}$ be an arbitrary symmetric invariant
tensor.  Then the order $m$ element in the enveloping algebra ${\cal
  U}(\g)$ defined by

\begin{equation}
  {\cal C}^{(m)}=h^{i_1\dots i_m} X_{i_1}\dots X_{i_m} \quad,
\end{equation}
commutes with all elements of $\g$. This is so because the commutator
$[X_\rho,{\cal C}^{(m)}]$ may be written as

\begin{equation}
  [X_\rho,{\cal C}^{(m)}]=\sum_{s=1}^m f_{\rho \nu } ^{i_s}
  h^{i_1\dots \widehat{i_s} \nu\dots  i_m} X_{i_1}\dots X_{i_m} =0 \quad,
  \label{invindices}
\end{equation}
which is indeed zero as a result of the invariance condition
(\ref{invcond}).  In fact, the only conditions for the $m$-tensor $h$
to generate a Casimir operator ${\cal C}^{(m)}$ of $\g$ of order $m$
are its symmetry (non-symmetric indices would allow us to reduce the
order $m$ of ${\cal C}^{(m)}$ by replacing certain products of
generators by commutators) and its invariance (eq.
(\ref{invindices})); $h$ does not need to be obtained from a symmetric
trace (\ref{symmtrace}).  Thus for any invariant symmetric tensor $h$
of order $m$, ${\cal C}^{(m)}=h^{i_1\dots i_m} X_{i_1}\dots X_{i_m}$
is a Casimir of $\g$ of the same order $m$.  
\vskip5pt 
\noindent It is well known
\cite{invGR,invRACAH,invGEL,invLCB,invAK,invPP,invOKPA,invSOK} that a
simple algebra of rank $l$ has $l$ independent (primitive)
Casimir-Racah operators of order $m_1,\dots,m_l$, the first of them
given by the familiar quadratic Casimir \cite{invCASIMIR} operator
$k_{ij}X^i X^j$ obtained from the Killing tensor $(m_1=2)$. Thus there
must be (Cayley-Hamilton) relations among the invariant tensors
obtained from (\ref{symmtrace}) for $m\neq m_l$ or otherwise one would
obtain an arbitrary number of primitive Casimirs.  We shall
demonstrate this explicitly for several examples.

\section{Relations among symmetric tensors}
\news
The results in this section involve an important result that, for a
matrix $A$, we have the relationship

\begin{equation}
  {\rm det} A = \exp (\tr \log A) \quad.
  \label{detAexpr}
\end{equation}
This is clearly true for diagonal $A$ and can be seen to be true
generally by conjugation of $A$. We apply this to the case $A =
{\bf 1}+\alpha F$, with ${\bf 1}$ the identity matrix. and $\alpha$ an
arbitrary parameter. We can treat the right hand side
of (\ref{detAexpr}) as a power series in $\alpha$, giving

\begin{eqnarray}
  \det ({\bf 1}+\alpha F) &=&1-\frac{\alpha^2}{2}\tr
  F^2+\frac{\alpha^3}{3}\tr F^3+\alpha^4\left[-\frac{1}{4}\tr
  F^4+\frac{1}{8}(\tr F^2)^2\right] \nonumber \\
  &+&\alpha^5\left[\frac{1}{5}\tr
  F^5-\frac{1}{6}\tr F^2
  \tr F^3\right] \nonumber \\
  &+&\alpha^6\left[-\frac{1}{6}\tr F^6+\frac{1}{18}(\tr
  F^3)^2+\frac{1}{8}\tr F^2 \tr F^4-\frac{1}{48}(\tr
  F^2)^3\right] \nonumber \\
  &+&\alpha^7\left[\frac{1}{7}\tr F^7-\frac{1}{10}\tr F^2
  \tr F^5-\frac{1}{12}\tr F^3 \tr F^4+\frac{1}{24}(\tr F^2)^2\tr
  F^3\right] \nonumber \\
  &+&\alpha^8\left[-\frac{1}{8}\tr F^8+\frac{1}{12}\tr
  F^2 \tr F^6+\frac{1}{15}\tr F^3 \tr F^5+\frac{1}{32} (\tr F^4)^2
  \right. \nonumber \\
  & &\left. -\frac{1}{32} \tr F^4 (\tr F^2)^2 -\frac{1}{36}(\tr F^3)^2
  \tr F^2+\frac{1}{16.4!} (\tr F^2)^4\right] + \ldots
  \label{detpowser}
\end{eqnarray}
In the above, $\tr F^n$ is defined as $F_{i_1 i_2}F_{i_2i_3} \cdots
F_{i_n i_1}$. All the terms to ${\cal O}(\alpha^8)$ are given as these are
the terms we shall use below. It is easy to calculate the terms in
higher powers of $\alpha$. We will now show how to express
non-primitive symmetric invariant tensors in terms of primitive tensors.

\subsection*{The algebras $A_l$}
We consider a matrix representation of $SU(n)$ with matrix generators
$\{X_i\}$. For our purposes we will always work with the fundamental
representation so that $X_i$ are $n \times n$ matrices. The rank of
the group is $l = n-1$ and its Lie algebra is $A_l =
\mathfrak{su}(l+1)$. A basis of the symmetric invariant tensors of
$A_l$ is given by the order $m$ symmetric traces $k^{(m)},
m=2,3,\ldots$ from equation (\ref{symmtrace}). The algebra $A_l$ has
$l$ primitive symmetric invariant tensors. A basis of these is given
by $k^{(m)}, m=2 \ldots n$. \vskip5pt

\noindent We now
define the matrix $F$ (see equation \ref{detpowser}) to be a linear
combination of the ${X_i}$, introducing a vector ${\bf y}$, as
$(F)_{ab} = y^i (X_i)_{ab}$. As $\{X_i\}$ are $n \times n$ matrices,
no terms of ${\cal O}(\alpha^m)$ can appear on the right hand side of
(\ref{detpowser}) for $m > n$. Thus these terms must vanish
identically, giving us one equation at each order in $\alpha$. This
gives us one equation for each of the non-primitive tensors of the
form (\ref{symmtrace}) in terms of primitive tensors.

\subsection*{The algebras $B_l$ and $C_l$}
The fundamental matrix representations of the algebras $B_l =
\mathfrak{so}(2l+1)$ and $C_l = \mathfrak{sp}(2l)$ are in terms of
$(2l+1)\times(2l+1)$ and $2l\times2l$ matrices respectively. In their
defining representations these groups preserve the euclidean or
symplectic metric $\eta$ respectively and thus the generators $X_i$ of
these algebras satisfy

\begin{equation}
  X_i \eta=-\eta X_i^t \quad,
  \label{preserveeta}
\end{equation}
where $i=1,\dots,l(2l+1)$. The symmetrised products of an {\it odd}
number of generators of the orthogonal and symplectic groups also
satisfies (\ref{preserveeta}); hence, they are a member of their
respective algebras. In particular, as the generators of both $B_l$
and $C_l$ are traceless, we find that the odd-order tensors
$k^{(2m+1)}$ (see eq.  \ref{symmtrace}), arising from symmetrized
traces, vanish identically. For both $B_l$ and $C_l$, we have the
property that a basis for the symmetric invariant tensors is given by
$k^{(m)}$ for $m=2,4,6,\ldots$. Thus for these examples, all odd
powers of $\alpha$ in the expansion (\ref{detpowser}) vanish
identically. A basis for the primitive symmetric invariant tensors is
given by $k^{(m)}, m=2,4,6, \ldots 2l$.  As above, we can write
$F_{ab} = y^i (X_i)_{ab}$, where $\{X_i\}$ is a basis for the required
Lie algebra in the fundamental representation.  Substituting this into
(\ref{detpowser}), we see that all terms of ${\cal O}(\alpha^m)$ must
vanish for $m>2l$. This gives precisely one equation for each of the
non-primitive tensors of the form (\ref{symmtrace}) in terms of
primitive tensors.

\subsection*{The algebras $D_l$} 
The algebra $D_l=\mathfrak{so}(2l)$ is the algebra of antisymmetric
$(2l)\times (2l)$ matrices. The odd-order tensors $k^{(2m+1)}$ vanish
identically using an identical argument to that used above for $B_l$.
There are $l$ primitive symmetric invariant tensors of which only
$(l-1)$ can be expressed in the form (\ref{symmtrace}). These are
$k^{(m)}, m=2,4,6,\ldots,2(l-1)$. The final primitive tensor is the
Pfaffian, given by

\begin{equation}
  Pf_{i_1i_2\ldots i_l} = \frac{1}{2^l l!} \epsilon^{a_1a_2\ldots a_{2l}}
    (X_{i_1})_{a_1a_2} (X_{i_2})_{a_3a_4} \ldots (X_{i_l})_{a_{(2l-1)}a_{2l}}
    .
\end{equation}
As before, we introduce a vector ${\bf y}$ and write $F_{ab} = y_i
(X^i)_{ab}$. We rewrite the expression (\ref{detpowser}) in the form

\begin{eqnarray}
  \det ({\bf 1}+\alpha F) &=&1+\alpha^2 \epsilon_{a_1a_2}^{b_1b_2}
    {F^{a_1}}_{b_1} {F^{a_2}}_{b_2} + \alpha^4
    \epsilon_{a_1a_2a_3a_4}^{b_1b_2b_3b_4} {F^{a_1}}_{b_1}
    {F^{a_2}}_{b_2} {F^{a_3}}_{b_3} {F^{a_4}}_{b_4} + \ldots \nonumber \\
  & + & \alpha^{2n} \epsilon_{a_1a_2\ldots a_{2n}}^{b_1b_2 \ldots b_{2n}}
    {F^{a_1}}_{b_1} {F^{a_2}}_{b_2} \ldots {F^{a_{2n}}}_{b_{2n}} +
    \ldots .
\end{eqnarray}
The terms of ${\cal O}(\alpha^{2m})$ for $m>l$ express
the non-primitive symmetric tensors $k^{(2m)}$ (see equation
\ref{symmtrace}) in terms of primitive tensors. For the ${\cal
  O}(\alpha^{2l})$ terms, we must use the identity

\begin{eqnarray}
  \epsilon_{a_1a_2\ldots a_{2l}}^{b_1b_2 \ldots b_{2l}}
    {F^{a_1}}_{b_1} {F^{a_2}}_{b_2} \ldots {F^{a_{2l}}}_{b_{2l}} & = &
    \left(\frac{1}{2^l l!} \epsilon_{a_1a_2\ldots
    a_{2l}}F_{a_1a_2} F_{a_3a_4}\ldots F_{a_{(2l-1)}a_{2l}} \right)^2
    \nonumber \\ 
  & = & Pf_{(i_1 \ldots i_l} Pf_{i_{l+1} \ldots i_{2l)}} y^{i_1} \ldots
    y^{i_{2l}} .
\end{eqnarray}
This expresses the non-primitive tensor $k^{(2l)}$ in terms of the
Pfaffian $Pf$ and the $(l-1)$ tensors $k^{(2m)}, m<l$, which together
are the $l$ primitive symmetric tensors of $D_l$.

\section{Examples}
\news

Below are given three explicit examples of construction of
non-primitive tensors for simple Lie algebras in terms of primitive
tensors. Connections are also made to the geometry of the associated
Lie groups. In particular, simple compact Lie groups can be expressed
as products of spheres of odd dimension, as far as the cohomology is
concerned. Decomposing a Lie group into such a product gives the de
Rham cohomology as the Betti numbers ~\cite{Betti} for a sphere $S^n$
are known to be

\begin{equation}
  b^{(0)}=1 \;, \quad b^{(n)}=1 \quad , \quad b^{(k)}=0 \quad 
    {\rm \mbox{otherwise}} \quad.
\end{equation}
We can also associate with each symmetric invariant tensor $k$ of order
$m$ an antisymmetric invariant tensor of order $(2m-1)$ according to

\begin{equation}
  \Omega_{i_1\dots i_{2m-2} \sigma} = f_{[i_1 i_2}^{j_1}\dots f_{i_{2m-3}
    i_{2m-2}}^{j_{m-1}} k_{\sigma] j_1\dots j_{m-1}} \quad.
  \label{cocycle}
\end{equation}
If $k$ is primitive, this gives a $(2m-1)$-form on $G$ representing a
Lie algebra cocycle of order $(2m-1)$. If $k$ is non-primitive, the
expression is identically zero ~\cite{deAMMandPB}.

\subsection*{The algebras $A_l$}
The group $SU(l+1)$ has algebra $\mathfrak{su}(l+1)=A_l$. The
cohomology of the group manifold is the cohomology of $l$ spheres as
$SU(l+1)\rightarrow S^{2l+1}\times S^{2l-1} \times \dots \times S^3$.
Thus there are $l$ non-trivial cocycles and $l$ corresponding
primitive symmetric invariant tensors as in (\ref{cocycle}). There is
a basis for the symmetric invariant tensors of $A_l$ given by Sudbery
~\cite{invSUDBERY}, consisting of one symmetric tensor $d^{(k)}$ at
each order $k=2,3,\ldots$. Given a normalized basis $\{\lambda_i\}$
for $\mathfrak{su}(l+1)$, we have the multiplication law

\begin{equation}
  \lambda_i \lambda_j =\frac{2}{l+1}\de_{ij}+\left(d_{ijk}+i f_{ijk}\right)\lambda_k\quad.
\end{equation}
The tensors $f$ and $d$ are respectively the totally antisymmetric and
symmetric structure constants. The two lowest-order symmetric
invariant tensors are $\de_{ij}$ and $d_{ijk}$. In terms of these, the
Sudbery basis for symmetric tensors is

\begin{eqnarray}
  d^{(4)}_{i_1i_2i_3i_4} &=&
    {d_{(i_1i_2}}^xd_{i_3i_4)x}\quad, \nonumber \\
  d^{(5)}_{i_1i_2i_3i_4i_5} &=&
    {d_{(i_1i_2}}^x{{d^x}_{i_3}}^y d_{i_4i_5)y}\quad, \\
  d^{(6)}_{i_1i_2i_3i_4i_5i_6} &=&
    {d_{(i_1i_2}}^x{{d^x}_{i_3}}^y{{d^y}_{i_4}}^z d_{i_5i_6)z}\quad.
    \nonumber \\
  &\vdots& \nonumber
\end{eqnarray}
We can choose a basis $d^{(k)}, k=2,3,\ldots,l+1$ of the primitive
  symmetric invariant tensors of $A_l$. These will give rise to
  representatives of the $l$ primitive cocycles of $A_l$ via
  (\ref{cocycle}).

\subsection*{The algebra $A_2$}
The fundamental representation of the algebra $A_2=\mathfrak{su}(3)$ is in
terms of eight $(3 \times 3)$ hermitian matrix generators
$\{\lambda_i\}$. The rank of $A_2$ is two so we have two primitive invariant
symmetric tensors; these are given by

\begin{equation}
  \frac{1}{2} \tr (\lambda_i \lambda_j)=\de_{ij} \quad,\quad
  \frac{1}{2}{\rm sTr}(\lambda_i \lambda_j \lambda_k)=d_{ijk} \quad.
  \label{SU3basis}
\end{equation}
As far as the cohomology is concerned, the group manifold $SU(3)$ can
be decomposed into $S^5 \times S^3$ and there are three- and
five-cocycles associated with $\de_{ij}$ and $d_{ijk}$ respectively
(see(\ref{cocycle})). Defining an arbitrary eight-vector ${\bf y}$,
writing $(F)_{ab}=y^i (X_i)_{ab}$ and substituting this into
(\ref{detpowser}) gives expressions for $d^{(m)}, m>3$ in terms of the
tensors (\ref{SU3basis}). Defining $k_{i_1\dots i_m}= {\rm
  sTr}(X_{i_1}\dots X_{i_m})$ as in (\ref{symmtrace}), the first few
of these are

\begin{eqnarray}
  k_{i_1i_2i_3i_4} &=&2 \de_{(i_1i_2} \de_{i_3i_4)}\quad, \\
  k_{i_1i_2i_3i_4i_5} &=&\frac{10}{3} \de_{(i_1i_2} d_{i_3i_4i_5)}\quad, \\
  k_{i_1i_2i_3i_4i_5i_6} &=&\frac{4}{3} d_{(i_1i_2i_3} d_{i_4i_5i_6)}+2\de_{(i_1i_2}\de_{i_3i_4}\de_{i_5i_6)}\quad.  
\end{eqnarray}
The first three non-primitive terms in the Sudbery basis for the
symmetric invariant tensors of $A_2$ can then be computed as

\begin{eqnarray}
  d^{(4)}_{i_1i_2i_3i_4} &=&
    {d_{(i_1i_2}}^xd_{i_3i_4)x}=\frac{1}{3}\de_{(i_1i_2}\de_{i_3)i_4}
    \quad, \nonumber \\
  d^{(5)}_{i_1i_2i_3i_4i_5} &=&
    {d_{(i_1i_2}}^x{{d^x}_{i_3}}^y d_{i_4i_5)y} =
    \frac{1}{3}\de_{(i_1i_2}d_{i_3i_4i_5)} \quad, \\
  d^{(6)}_{i_1i_2i_3i_4i_5i_6} &=&
    {d_{(i_1i_2}}^x{{d^x}_{i_3}}^y{{d^y}_{i_4}}^z d_{i_5i_6)z} =
    \frac{2}{15}d_{(i_1i_2i_3}d_{i_4i_5i_6)}+\frac{1}{15}\de_{(i_1i_2}\de_{i_3i_4}\de_{i_5i_6)} \quad. \nonumber \\ 
\end{eqnarray}

\subsection*{The algebras $B_n$}
The fundamental representation of the algebra
$B_n=\mathfrak{so}(2n+1)$ is in terms of $(2n+1)\times(2n+1)$
antisymmetric matrices. We define a basis $\{X_i\}$, normalized as
$\tr(X_i X_j)=2\de_{ij}$.  There will be $n$ primitive symmetric
invariant tensors (equal to the rank of the group). The first two of
these are $\delta_{ij} = \frac{1}{2}\tr(X_i X_j)$ and the order four
totally symmetric tensor ~\cite{deAzcarragaandPB} given by

\begin{equation}
  \tr (X_{(i} X_j X_{k)}) = d^{(4)}_{ijkl} X_l \quad.
\end{equation}
We notice that the tensors $k^{(m)}$ (given by (\ref{symmtrace}))
vanish for $m$ odd and we have a basis for the symmetric tensors
of $B_n$ in analogy with Sudbery's basis for $A_n$, {\it i.e.}

\begin{eqnarray}
  d^{(6)}_{i_1i_2i_3i_4i_5i_6} &=&
    {d_{(i_1i_2i_3}}^xd_{i_4i_5i_6)x} \quad, \nonumber \\
  d^{(8)}_{i_1i_2i_3i_4i_5i_6i_7i_8} &=&
    {d_{(i_1i_2i_3}}^x{{d^x}_{i_4i_5}}^y{d^y}_{i_6i_7i_8)}\quad. \\
  &\vdots& \nonumber
\end{eqnarray}
The algebra $B_n$ will have n primitive symmetric tensors of orders
$2,4,\dots,2n$, giving rise to cocycles of orders $3,7,\dots,(4n-1)$
via (\ref{cocycle}).

\subsection*{The algebra $B_2$}
The fundamental representation of the algebra $B_2=\mathfrak{so}(5)$
is in terms of $(5\times5)$ antisymmetric matrices with a basis
$\{X_i\}$, normalized as above. As far as the cohomology is concerned,
the group manifold decomposes as $SO(5) \sim S^3 \times S^7$, giving
non-trivial three- and seven-cocycles. The three-cocycle is associated
with the symmetric tensor $\delta_{ij} = \frac{1}{2}\tr(X_i X_j)$ and
the seven-cocycle with the order four symmetric tensor
$d^{(4)}_{ijkl}=\frac{1}{2} \tr (X_{(i} X_j X_{k)} X_l)$ according to
(\ref{cocycle}). As above, we define an arbitrary vector ${\bf y}$,
write $(F)_{ab}=y^i (X_i)_{ab}$ and substitute this into
(\ref{detpowser}). The antisymmetry of the algebra elements means that
all traces of odd powers of $F$ vanish identically, giving the result
at ${\cal O}(\alpha^6)$:

\begin{equation}
  \tr F^6 = \frac{3}{4}\tr F^2\tr F^4 - \frac{1}{8}(\tr F^2)^3 \quad,
\end{equation}
or equivalently

\begin{equation}
  k^{(6)}_{i_1i_2i_3i_4i_5i_6}=\frac{3}{2}\de_{(i_1i_2}k^{(4)}_{i_3i_4i_5i_6)} - \de_{(i_1i_2}\de_{i_3i_4}\de_{i_5i_6)} \quad.
\end{equation}
Expressions for the higher-order symmetric tensors can be obtained easily.

\subsection*{The algebra $G_2$}

The algebra $G_2$ can be embedded into $\mathfrak{so}(7)$; the
generators of $SO(7)$ decompose into representations ${\bf 21}
\rightarrow {\bf 7} \oplus {\bf 14}$ under the action of $G_2$. We
will work with the seven-dimensional representation of $G_2$ embedded
into the fundamental representation of $\mathfrak{so}(7)$. This has
generators $\{X_i\}$, normalized as $\tr(X_i X_j)=2\de_{ij}$. As for
the orthogonal groups, the terms odd in $F$ in the expansion
(\ref{detpowser}) will vanish identically.  It is known
~\cite{deAzcarragaandPB} that there are two primitive symmetric
invariant tensors of $G_2$, of order two and six, given by the tensors
$k^{(2)}$ and $k^{(6)}$ defined in (\ref{symmtrace}). We note the result
from ~\cite{Okubo}:

\begin{equation}
  {\rm sTr}(X_i X_j X_k X_l)=\frac{1}{4}\tr(X_{(i} X_j)\tr(X_k
    X_{l)}) \quad.
\end{equation}
Using this and the methods described above, we can now write any
symmetric invariant tensor of $G_2$ in terms of the two primitive
symmetric tensors

\begin{equation}
  k_{i_1i_2}=\tr(X_{i_1}X_{i_2})\quad,\quad
  k_{i_1i_2i_3i_4i_5i_6}={\rm sTr}(X_{i_1}X_{i_2}X_{i_3}X_{i_4}X_{i_5}X_{i_6})\quad.
\end{equation}

\section{Conclusion}
A simple Lie group has a number of primitive symmetric invariant
tensors equal to the rank of the group. Other symmetric invariant
tensors must be expressible in terms of these. This is equivalent to
expressing non-primitive Casimirs in terms of primitive Casimirs. We
have presented a systematic method for doing this which has been
illustrated with three examples including one example of an
exceptional Lie algebra.

\section*{Acknowledgements}
I thank Alan Macfarlane, Jonathan Evans and Niall Mackay for
interesting and helpful discussions. I acknowledge the financial
assistance of a PPARC studentship.


\newpage

\begin{thebibliography}{99}

\bibitem{invGR}
B.~Gruber and L.~O'Raifeartaigh.
\newblock $s$-theorem and construction of the invariants of the semisimple
  compact {L}ie algebras.
\newblock {\em J. Math. Phys.}, 5:1796--1804, 1964.

\bibitem{invRACAH}{G. Racah,
{\it Sulla caratterizzazione delle rappresentazioni irreducibili dei gruppi
semisimplici di Lie},
Lincei-Rend. Sc. fis. mat. e nat. VIII, 108-112, 1950;
Princeton lectures, CERN-61-8 (reprinted in Ergeb. Exact Naturwiss.
37, 28-84, 1965), Springer-Verlag}

\bibitem{invGEL}
{I.M. Gel'fand, {\it The center of an infinitesimal group ring},
Math. Sbornik 26, 103-112, 1950. (English transl.: Los Alamos
Sci. Lab. AEC-TR-6133, 1963 )}

\bibitem{invLCB}
L.~C. Biedenharn.
\newblock On the representations of the semisimple {L}ie groups {I}.
\newblock {\em J. Math. Phys.}, 4:436--445, 1963.

\bibitem{invAK}
A.~Klein.
\newblock Invariant operators of the unimodular group in $n$ dimensions.
\newblock {\em J. Math. Phys.}, 4:1283--1284, 1963.

\bibitem{invPP}
A.~M. Perelomov and V.~S. Popov.
\newblock Casimir operators for semisimple groups.
\newblock {\em Math. USSR-Izvestija}, 2:1313--1335, 1968.

\bibitem{invOKPA}
S.~Okubo and J.~Patera.
\newblock General indices of representations and {C}asimir invariants.
\newblock {\em J. Math. Phys.}, 25:219--227, 1983.

\bibitem{invSOK}
S.~Okubo.
\newblock Modified fourth-order {C}asimir invariants and indices for simple
  {L}ie algebras.
\newblock {\em J. Math. Phys.}, 23:8--20, 1982.

\bibitem{invCASIMIR}
H.B.G. Casimir.
\newblock {\em Proc. Roy. Acad. Amsterdam}, 34:844, 1931.

\bibitem{Betti}
E.~Betti.
\newblock {\em Ann. Mat. Pura Appl.}, 4:140--158, 1871.

\bibitem{deAMMandPB}
J.~A. de~Azc\'arraga, A.~J. Macfarlane, A.~J. Mountain, and J.~C.~P\'erez
  Bueno.
\newblock Invariant tensors for simple groups.
\newblock {\em Nucl. Phys.}, B510:657--687, 1998.

\bibitem{invSUDBERY}
{A. Sudbery, Ph.D. Thesis, Cambridge Univ., 1970;
{\it Computer-friendly $d$-tensor identities for $SU(n)$}
J. Phys. A23(15), L705-L710, 1990}

\bibitem{deAzcarragaandPB}
J.A. de~Azc\'arraga and J.C.~P\'erez Bueno.
\newblock Higher order simple {L}ie algebras.
\newblock {\em Commun. Math. Phys.}, 184:669--681, 1997.

\bibitem{Okubo}
S.~Okubo.
\newblock Casimir invariants and vector operators in simple and classical {L}ie
  algebras.
\newblock {\em J. Math. Phys.}, 18(12):2382--2394, 1977.

\end{thebibliography}
\end{document}